\documentstyle[12pt]{article}
\input epsf
\def\whatjournal{N}
\newlinechar=`\^^J

\def\ordernpb#1#2#3{{\bf#1} (#3) #2}
\if P\whatjournal {\global\def\order#1#2#3{\orderprd{#1}{#2}{#3}}}
                    \immediate\write16{^^J PRD references ^^J}\else
                   {\global\def\order#1#2#3{\ordernpb{#1}{#2}{#3}}}
                    \immediate\write16{^^J NPB references ^^J}
\fi
\def\ijmpa#1#2#3{{\it Int. J. of Mod. Phys. {\bf A}}\order{#1}{#2}{#3}}

\def\npb#1#2#3{{\it Nucl. Phys. {\bf B}}\order{#1}{#2}{#3}}
\def\plb#1#2#3{{\it Phys. Lett. {\bf B}}\order{#1}{#2}{#3}}

\def\prep#1#2#3{{\it Phys. Rep.\ }\order{#1}{#2}{#3}}

\def\prd#1#2#3{{\it Phys. Rev. {\bf D}}\order{#1}{#2}{#3}}

\def\rmp#1#2#3{{\it Rev. Mod. Phys.\ }\order{#1}{#2}{#3}}
\def\zphys#1#2#3{{\it Z. Phys. {\bf C}}\order{#1}{#2}{#3}}

\def\and{{\it\&}}
\def\half{{1\over2}}

\def\quarter{{1\over4}}

\def\gesim{\,{\raise-3pt\hbox{$\sim$}}\!\!\!\!\!{\raise2pt\hbox{$>$}}\,}
\def\lesim{\,{\raise-3pt\hbox{$\sim$}}\!\!\!\!\!{\raise2pt\hbox{$<$}}\,}
\def\boldoverdot{\,{\raise6pt\hbox{\bf.}\!\!\!\!\>}}

\def\etal{{\it et. al.}}
\def\acal{{\cal A}}

\def\lcal{{\cal L}}

\def\ncal{{\cal N}}
\def\ocal{{\cal O}}

\def\vev{vacuum expectation value}

\def\diag{\hbox{\diag}}

\def\gev{\hbox{GeV}}
\def\tev{\hbox{TeV}}

\def\inbox#1{\vbox{\hrule\hbox{\vrule\kern5pt
     \vbox{\kern5pt#1\kern5pt}\kern5pt\vrule}\hrule}}
\def\sqr#1#2{{\vcenter{\hrule height.#2pt
      \hbox{\vrule width.#2pt height#1pt \kern#1pt
         \vrule width.#2pt}
      \hrule height.#2pt}}}

\def\today{\ifcase\month\or
  January\or February\or March\or April\or May\or June\or
  July\or August\or September\or October\or November\or December\fi
  \space\number\day, \number\year}
\def\pmb#1{\setbox0=\hbox{#1}%
  \kern-.025em\copy0\kern-\wd0
  \kern.05em\copy0\kern-\wd0
  \kern-.025em\raise.0433em\box0 }
\def\up#1{^{\left( #1 \right) }}
\def\lowti#1{_{{\rm #1 }}}
\def\inv#1{{1\over#1}}
\def\su#1{{SU(#1)}}
\def\ui{U(1)}
\def\sumprime_#1{\setbox0=\hbox{$\scriptstyle{#1}$}
  \setbox2=\hbox{$\displaystyle{\sum}$}
  \setbox4=\hbox{${}'\mathsurround=0pt$}
  \dimen0=.5\wd0 \advance\dimen0 by-.5\wd2
  \ifdim\dimen0>0pt
  \ifdim\dimen0>\wd4 \kern\wd4 \else\kern\dimen0\fi\fi
\mathop{{\sum}'}_{\kern-\wd4 #1}}
\font\smalli=cmr8 scaled\magstep1
\font\smallii=cmr8 scaled\magstep2

\def\sm{Standard Model}
\def\ltv{\Lambda\lowti{TeV}}
\def\thecaption#1#2{\centerline{\vbox to 1 in{\hsize 5 in \vfill
           {\textindent{#1} \global \advance \baselineskip by -10 pt
          \smalli \noindent#2 }}}\global \advance \baselineskip by 10 pt}
\def\mz{ m\lowti{z} }
\def\mw{ m\lowti{w} }
\def\mh{ m\lowti{H} }
\def\sw{ s\lowti{w} }
\def\cw{ c\lowti{w} }
\def\cww{ c\lowti{ 2 w } }
\def\nsd{\ncal\lowti{SD}}
\def\thepreprintnumber{IFT-2/95, UCR-T133}

\def\putpreprintnumber{{\smallii \\ \vskip -115 pt \hfill \thepreprintnumber
\\ \vskip -8 pt\hfill \today \vskip 123 pt} }

\title{  Higgs boson production
at $ e^+ e^- $ colliders: a model independent approach.
\putpreprintnumber
}

\author{
Bohdan Grz\c{a}dkowski
\vspace*{0.6em}\\
{\normalsize \sl Department of Physics}\\
{\normalsize \sl Warsaw University}\\
{\normalsize \sl Warsaw, PL-00-681 Poland} \\
\vspace*{1.2em}\\
Jos\'{e} Wudka
\vspace*{0.6em}\\
{\normalsize \sl Department of Physics}\\
{\normalsize \sl University of California, Riverside}\\
{\normalsize \sl Riverside, CA 92521-0413 U.S.A.}
}
\date{\today}

\begin{document}

\maketitle

\vfil

\begin{abstract}
\rm
We consider non-\sm\ physics effects using an effective lagrangian
parameterization. We determine the operators whose effects are most
significant and extract the sensitivity to the scale of new physics
generated by the existing data. We then consider processes containing
the Higgs particle in $ e^+e^-$ colliders
as a probe for new physics effects, and demonstrate
their usefulness in this area.
\end{abstract}

\pagebreak

\renewcommand{\arraystretch}{1.5}
\renewcommand{\textheight}{7.51in}

The possibility of determining new physics effects by precision
measurements has been pursued for a long time in an effort to provide
insights into the interactions that lie beyond the \sm. Many of these
efforts were carried out for a specific model of non-\sm\
physics~\cite{Langacker}.

Recently it has become evident that these studies should be
complemented with a model independent approach where all possible
non-standard effects are parameterized by means of an effective
lagrangian~\cite{leffref,gi}. This formalism is model and process independent
and
thus
provides an unprejudiced analysis of the data. Such an approach will be
pursued in this paper. We will parameterize all non-standard effects using
the coefficients of a set of effective operators (which respect the
symmetries of the \sm). These operators are
chosen so that there are no {\it a-priori}
reasons to suppose that the said coefficients are suppressed. In this
respect the present analysis differs from others appearing in the
literature~\cite{others} which concentrate on operators related to the
vector-boson self interactions. In using a manifestly gauge invariant
parametrization we diverge from those studies aimed at
elucidating the rigidity of the \sm\ to violations of its symmetries
(whose drawbacks have been emphasized previously~\cite{gi}).

The effective lagrangian approach requires a choice of the low energy
particle content. In this paper we will assume that the \sm\ correctly
describes all such excitations (including the Higgs particle)~\footnote{Other
approaches can be followed, assuming, for example, an extended scalar
sector or the complete absence of light physical scalars.}. Thus we
imagine that there is a scale $ \Lambda $, independent of the Fermi
scale, at which the new physics becomes apparent. Since the \sm\ is
renormalizable and the new physics is assumed to be heavy due to a large
dimensional parameter $ \sim \Lambda $, the decoupling
theorem~\cite{Appelquist} is applicable and requires
that all new physics effects be suppressed by inverse powers of $
\Lambda $. All such effects are expressed in terms of a series of local
{\it gauge invariant}
operators of canonical dimension $ > 4 $; the catalogue of such
operators up to dimension 6 is given in Refs.~\cite{bw} (there are no
dimension 5 operators respecting the global and local symmetries of the
\sm).

For the situation we are considering it is natural to assume that the
underlying theory is weakly coupled (else radiative corrections will
drive the Higgs mass to $ \Lambda $ unless the low energy particle
content is modified to effect cancelations, we will not pursue this
possibility). {\it Thus the relevant property of a given
dimension 6 operator is whether it can be generated at tree level by
the underlying physics.} The coefficient of such operators are expected
to be $ O ( 1 ) $; in contrast, the coefficients of loop-generated
operators will contain a suppression~\footnote{If there is a
large number of loop graphs this suppression
factor can be reduced but, simultaneously, the masses not
protected by a symmetry will, in general, be driven to the scale $
\Lambda $.} factor
$ \sim 1 / 16 \pi^2 $.
The determination of those operators which are tree level generated is
given in Ref.~\cite{Arzt.et.al.}.

The strategy which we follow in this paper is to develop the effects of
the tree-level-generated operators containing leptons and scalars in
various processes. We will consider the constraints implied by current
high-precision data and predict the sensitivity to new effects
at LEP2 and a proposed version of the NLC.

The operators which we will consider are
\begin{equation}
\begin{array}[c]{ll}
\ocal_\phi = \inv3 \left( \phi^\dagger \phi \right)^3 &
\ocal_{ \partial\phi} = \half \partial_\mu \left( \phi^\dagger \phi
                          \right) \; \partial^\mu \left( \phi^\dagger \phi
                          \right) \\
\ocal\up1_\phi = \left( \phi^\dagger \phi \right) \left[ \left( D_\mu
                   \phi \right)^\dagger  D^\mu \phi \right] &
\ocal\up3_\phi = \left( \phi^\dagger D_\mu \phi \right) \left[ \left(
                    D_\mu \phi \right)^\dagger  \phi \right] \\
\ocal\up1_{ \phi \ell } = i \left( \phi^\dagger D_\mu \phi \right)
                           \left( \bar \ell \gamma^\mu \ell \right) &
\ocal\up3_{ \phi \ell } = i \left( \phi^\dagger \tau^I D_\mu \phi \right)
                             \left( \bar \ell \tau^I \gamma^\mu \ell\right) \\
\ocal_{ \phi e } = i \left( \phi^\dagger D_\mu \phi \right)
                           \left( \bar e \gamma^\mu e \right) & \\
\end{array} \label{eq:operators}
\end{equation}
Where we have omitted those that contain quark fields. The complete list
of tree level generated operators can be found in Ref.~\cite{Arzt.et.al.}.
Note that the modifications to the $WWZ$ and $WW\gamma$ vertices are not
in this list; this implies, as has been repeatedly emphasized, that
LEP2 will not be sensitive to these effects. The NLC will have enough
sensitivity to probe these anomalous couplings, still its sensitivity to
non-\sm\ processes generated by (\ref{eq:operators}) will be
significantly larger. {\it The effects of the above operators present the
widest windows into physics beyond the \sm.}

Given the above list the lagrangian which we will use in the following
calculations is
\begin{equation}
\lcal = \lcal\up{SM} + \inv{ \Lambda^2 } \sum_i\left\{ \alpha_i \ocal_i +
          \hbox{ h.c.} \right\} \label{eq:lagrangian}
\end{equation}

The above operators modify the couplings of the leptons to the $Z$ and
to the $W$ gauge bosons; they also modify the $ \rho $ parameter, the
Fermi constant and the normalization of the Higgs field. Only this last
effect is not probed in the existing data.

The electron vector and axial couplings to the $Z$, $ g_V(e) $ and $
g_A(e) $, and the neutrino coupling to the $Z$, $ g_\nu $, receive the
contributions~\cite{opal}
\begin{eqnarray}
\left| \delta g_V (e) \right| &=&
{ v^2 \over 2 \Lambda^2 } \left| \alpha_{ \phi \ell }
\up1 + \alpha_{ \phi \ell }\up3 + \alpha_{ \phi e } \right|
\lesim 0.0021 , \nonumber \\
\left| \delta g_A(e) \right| &=&
{ v^2 \over 2 \Lambda^2 } \left| \alpha_{ \phi \ell }
\up1 + \alpha_{ \phi \ell }\up3 - \alpha_{ \phi e } \right| \nonumber
\lesim0.00064 , \\
\left| \delta g_\nu \right| &=&
{ v^2 \over 2 \Lambda^2 } \left| \alpha_{ \phi \ell }
\up1 - \alpha_{ \phi \ell }\up3 \right| \lesim 0.0018 ; \nonumber \\
\end{eqnarray}
where the ($ 1 \sigma $) experimental constraints are also indicated.
At the $ 3 \sigma $ level these bounds then correspond to the constraints
\begin{equation}
\ltv \gesim { 2.5 \over \sqrt{ \left| \alpha_{ \phi \ell } \up1 \right| } }
, \; { 2.5 \over \sqrt{ \left| \alpha_{ \phi \ell } \up3 \right| } } , \;
{ 2.7 \over \sqrt{ \left| \alpha_{ \phi
e } \right|} }, \label{eq:lepbounds}
\end{equation}
where $ \ltv $ is the scale of new physics in \tev\ units.

Similarly the contributions to the $ \rho $ parameter arise form $
\ocal_\phi\up3 $, explicitly
\begin{equation}
\left| \delta T  \right| = { 4 \pi \over \sw^2 } \left| \alpha_\phi \up 3
\right|
{ v^2 \over \Lambda ^2 } \lesim 0.4
\label{eq:tbounds} \end{equation}
This bound~\cite{pdb} implies $ \ltv \gesim 1.7/ \sqrt{
\left| \alpha_\phi \up 3 \right| } $ (at $ 3 \sigma $).

The Fermi constant receives contributions from $ \ocal_{ \phi
\ell } \up 3 $, $ \ocal_\phi\up1 $ and form the four fermion operator $
\left( \bar \ell \gamma_\mu \tau^I \ell \right)^2 /2 $. {\it We will use $
G_F $, the fine structure constant and the $Z$ mass as our input
parameters}, then the effects of the $ \ocal $ on $G_F$ and $ m_Z $ are
observed in deviations of the $W$ mass form its \sm\ prediction. None of
the high precision measurements constrain $ \alpha_\phi\up1 $ since,
without direct observation of the Higgs, the tree-level effects of this
operator are absorbed in the wave function renormalization of the scalar
doublet.

The excellence of the above constraints has been used to claim that the
possibilities of observing new physics effects at LEP2 are greatly
diminished (if not absent), having been preempted by
LEP1~\cite{Derujula}. While this is true for
all effects relating to the coupling of the fermions to the gauge bosons,
it is not so for the couplings of the Higgs to the other fields. It
is precisely on these effects that we will concentrate upon in the
following. It will be assumed that the Higgs will be detected and
studied and we will use the high precision measurements expected from LEP2
to constrain physics beyond the \sm\ by observation of reactions where
the Higgs is produced.

We will concentrate on two reactions, namely
\begin{equation}
e^+ e^- \rightarrow Z H \qquad e^+ e^- \rightarrow H \bar \nu \nu
\end{equation}
At LEP2 the second reaction is dominated by the Bjorken process
(followed by $Z\rightarrow \bar \nu \nu$), at
higher energies $W$ fusion will dominate (at least within the $ R_\xi $
gauges); we will include both sets of graphs.

Using the expression (\ref{eq:lagrangian}), with $ \ocal_i $ defined in
(\ref{eq:operators}), we calculate the relevant amplitude for $
e^+ e^- \rightarrow Z H $; the Feynman graphs are given in figure 1. The
Feynman rules are extracted as usual from (\ref{eq:lagrangian}), the
resulting amplitude squared is
\begin{equation}
\left| \acal \right|^2  = \quarter \left( \left| a_L \right|^2
+ \left| a_R \right|^2 \right) \left[ s + { ( t - \mz^2 )( u - \mz^2 )
\over \mz^2 } \right]
\end{equation}
where
\begin{eqnarray}
a_L &=& { 2 g \over v \cw } { \mz^2 \over s - \mz^2 } \left[ { 2 \sw^2 -
1 \over 2 } \left( 1 + \delta_Z \right) + \delta \epsilon_L { s
\over \mz^2 } \right]
\nonumber \\
a_R &=& { 2 g \over v \cw } { \mz^2 \over s - \mz^2 } \left[ \sw^2
\left( 1 + \delta_Z \right) + \delta \epsilon_R { s
\over \mz^2 } \right] \nonumber \\
\end{eqnarray}
and
\begin{eqnarray}
\delta_Z &=& { v^2 \over 2 \Lambda^2 } \left(  \alpha_\phi\up1 +
\alpha_\phi\up3 - 2 \alpha_{ \partial \phi } \right) \nonumber \\
\delta \epsilon_L &=& - { v^2\over 2 \Lambda^2 }  \left(
\alpha_{ \phi \ell } \up1 +  \alpha_{ \phi \ell } \up3 \right)
\nonumber \\
\delta \epsilon_R &=& - { v^2\over 2 \Lambda^2 } \alpha_{ \phi e }
\nonumber \\
\end{eqnarray}

\noindent where $s$, $t$ and $u$ stand for the usual Mandalstam variables.
As mentioned above, we
will use the Fermi constant $ G_F$, the $Z$ mass $ \mz $ and the fine
structure constant  as input parameters; all other quantities are to be
expressed in terms of these numbers. We used $ \cw =
g/ \sqrt{ g^2 + g' {}^2 } $ and $ \sw =
g'/ \sqrt{ g^2 + g' {}^2 } $; the
\vev\ $v$ denotes the minimum of the scalar potential and receives a
modification due to the presence of $ \ocal_\phi $.
The mass $ \mz $ denotes the physical mass of the $Z$
particle.

\bigskip
\setbox2=\vbox to .8 truein{\epsfysize=4 truein\epsfbox[0 0 612 792]{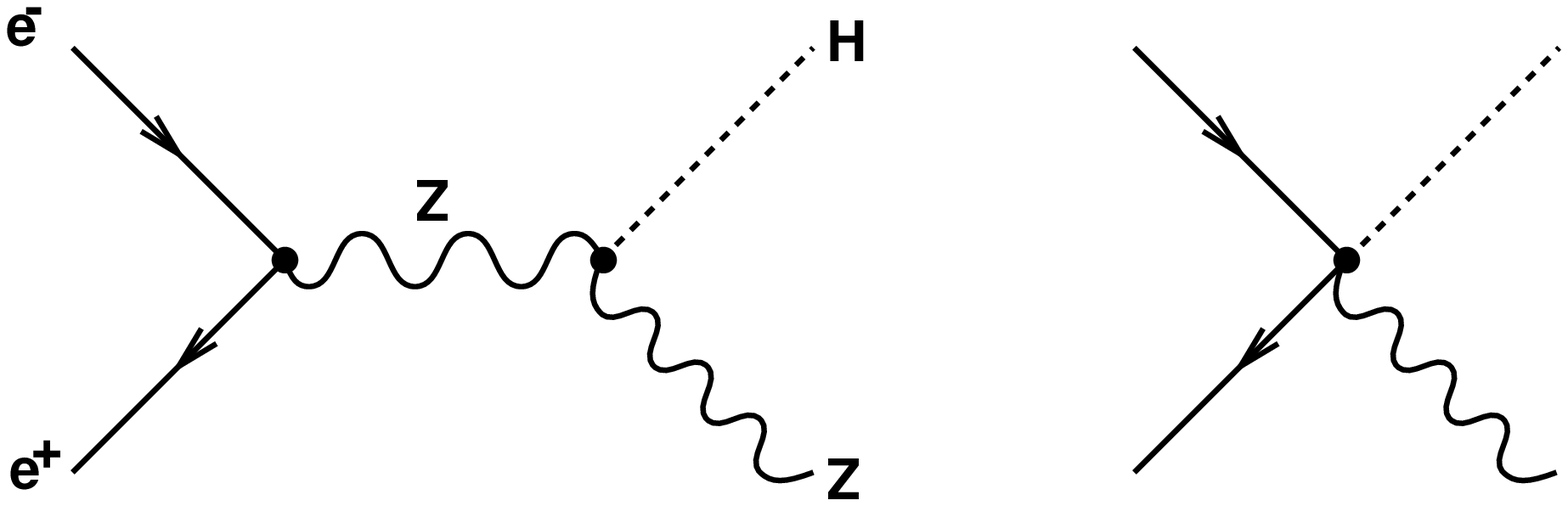}}
\centerline{\box2}
\thecaption{Fig. 1}{Feynman graphs contributing to the process $ e^+ e^-
\rightarrow Z H $.}
\bigskip\bigskip

Similarly we evaluate the amplitude for the process $ e^+ e^-
\rightarrow H \bar \nu \nu $. The Feynman graphs are given in figure 2. The
resulting amplitude squared is
\begin{equation}
\left| \acal \right|^2 = 8 | X_v |^2 \left( p_2' \cdot p_1
\right) \left( p_2 \cdot p_1' \right) + 2 | X_s |^2 \left( p_2' \cdot p_1'
\right) \left( p_2 \cdot p_1 \right)
\end{equation}
where
\begin{eqnarray}
X_v &=& { g^2 \over v } { \mw^2 \over ( q_1^2 - \mw^2 ) ( q_2^2 - \mw^2 )
} \left[ 1 + \delta_W + { q_1^2 + q_2^2 \over \mw^2 } \delta g_L \right]
\nonumber \\
&& \quad + { g^2 - g'{}^2 \over 2 v } { \mz^2 \over ( k_1^2 - \mz^2 ) (
k_2 - \mz^2 ) } \left[ 1 + \delta_Z + { 2 \delta_{ \nu_L } k_2^2 -
( 2 / \cww ) \delta \epsilon_L k_1^2 \over \mz^2 } \right]
\nonumber \\
X_s &=& { 2 g'{}^2 \over v } { \mz^2 \over ( k_1^2 - \mz^2 ) (
k_2 - \mz^2 ) } \left[ 1 + \delta_Z + { 2 \delta_{ \nu_L } k_2^2 +
( 1 / \sw^2 ) \delta \epsilon_L k_1^2 \over \mz^2 } \right]
\nonumber \\
\end{eqnarray}

In the above
\begin{eqnarray}
\delta g_L &=& { v^2 \over \Lambda^2 } \alpha_{ \phi \ell } \up 3
\nonumber \\
\delta_{ \nu_L } &=& { v^2 \over 2 \Lambda^2 } \left(
\alpha_{ \phi \ell } \up 3 - \alpha_{ \phi \ell } \up 1 \right)
\nonumber \\
\delta_W &=& - { v^2 \over 2 \Lambda^2 }  \left(
2 \alpha_{ \partial \phi } - \alpha_\phi\up1 + \alpha_\phi\up3
\right) ; \nonumber \\
\end{eqnarray}
the other quantities were defined previously.
The momentum assignments are indicated in figure 2; $ p_{ 1 , 2 } $ are
incoming while $ p'_{ 1 , 2 } $ are outgoing.

\bigskip\bigskip
\setbox2=\vbox to 1.7 truein{\epsfysize=4 truein\epsfbox[0 0 612 792]{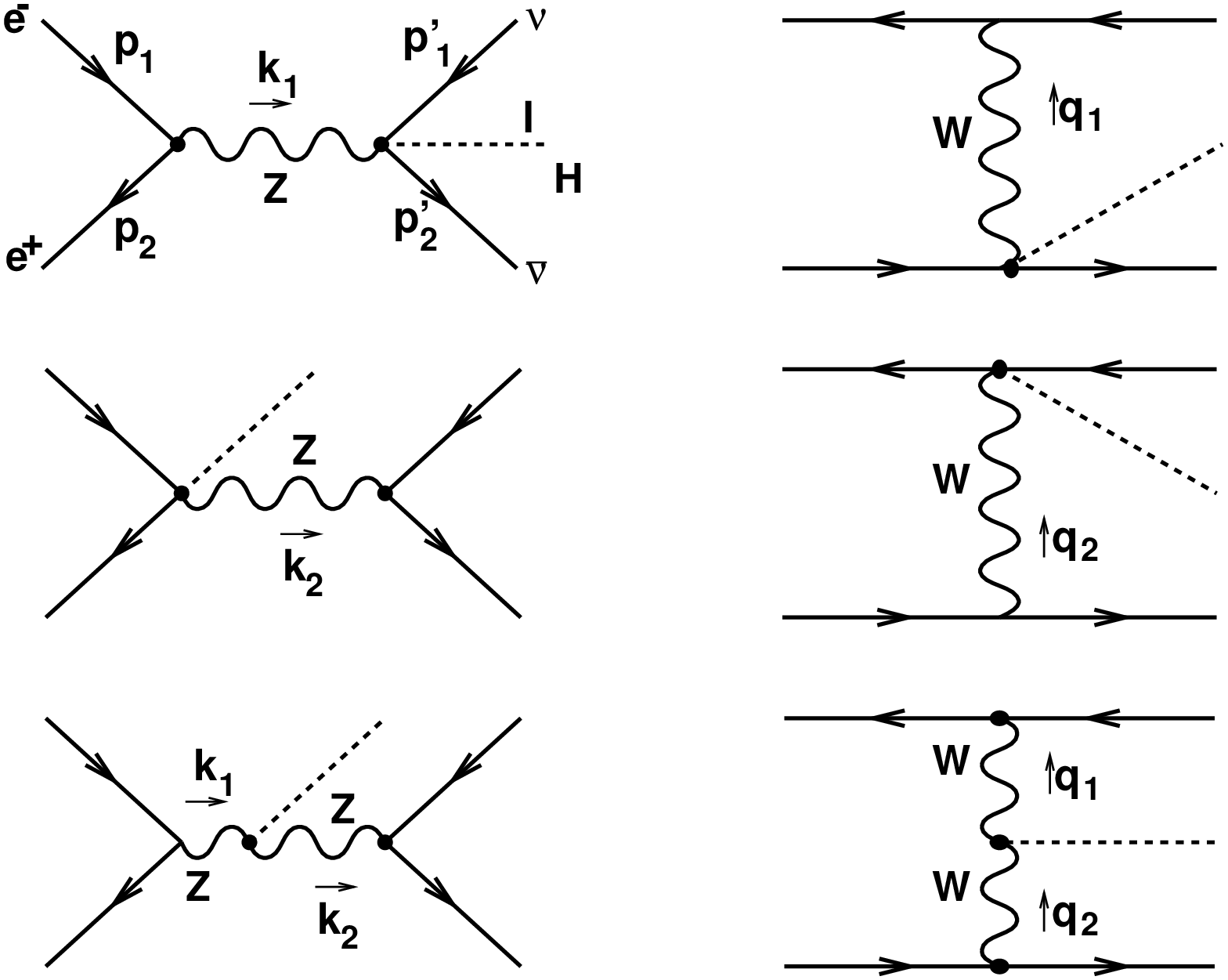}}
\centerline{\box2}
\thecaption{Fig. 2}{Feynman graphs contributing to the process $ e^+ e^-
\rightarrow H \bar \nu \nu $.}
\bigskip\bigskip

Note that the corrections to the \sm\ expressions are of two types. There
are small modifications to the coefficients and, more importantly, there
are terms which grow with the momenta and which appear to violate
unitarity for sufficiently large $s$. This is, of course, only a signal
that the approximations used break down for sufficiently large energy;
this problem is never encountered since (\ref{eq:lagrangian}) is only
valid for $ s \ll \Lambda^2 $. Nonetheless the presence of these terms
generates the most significant deviations from the \sm. This effect is
reminiscent of the delayed unitarity effects~\cite{Ahn et.al.}
which, though not dramatic,
do provide non-trivial sensitivity into new physics effects.

The results for the total cross sections for the two processes are given
in figure 3. In presenting these results we chose values for
$ \alpha_{\phi \ell } \up{ 1 , 3 }  $, $ \alpha_{\phi e } $ and $
\alpha_\phi \up3 $ which
saturate the ($ 3 \sigma $) bounds derived from
(\ref{eq:lepbounds},\ref{eq:tbounds}); the signs were chosen to
maximize the effects. The rest of the $ \alpha_i $ were chosen equal to
one. We also restricted the
range of $s$ to the point where the deviations from the \sm\  generated
by the dimension 6 operators are $ \sim $100\%,
at which point the approximations used break down.

\bigskip\bigskip
\setbox2=\vbox to 200 pt{\epsfxsize=6 truein\epsfbox[0 -140 612
652]{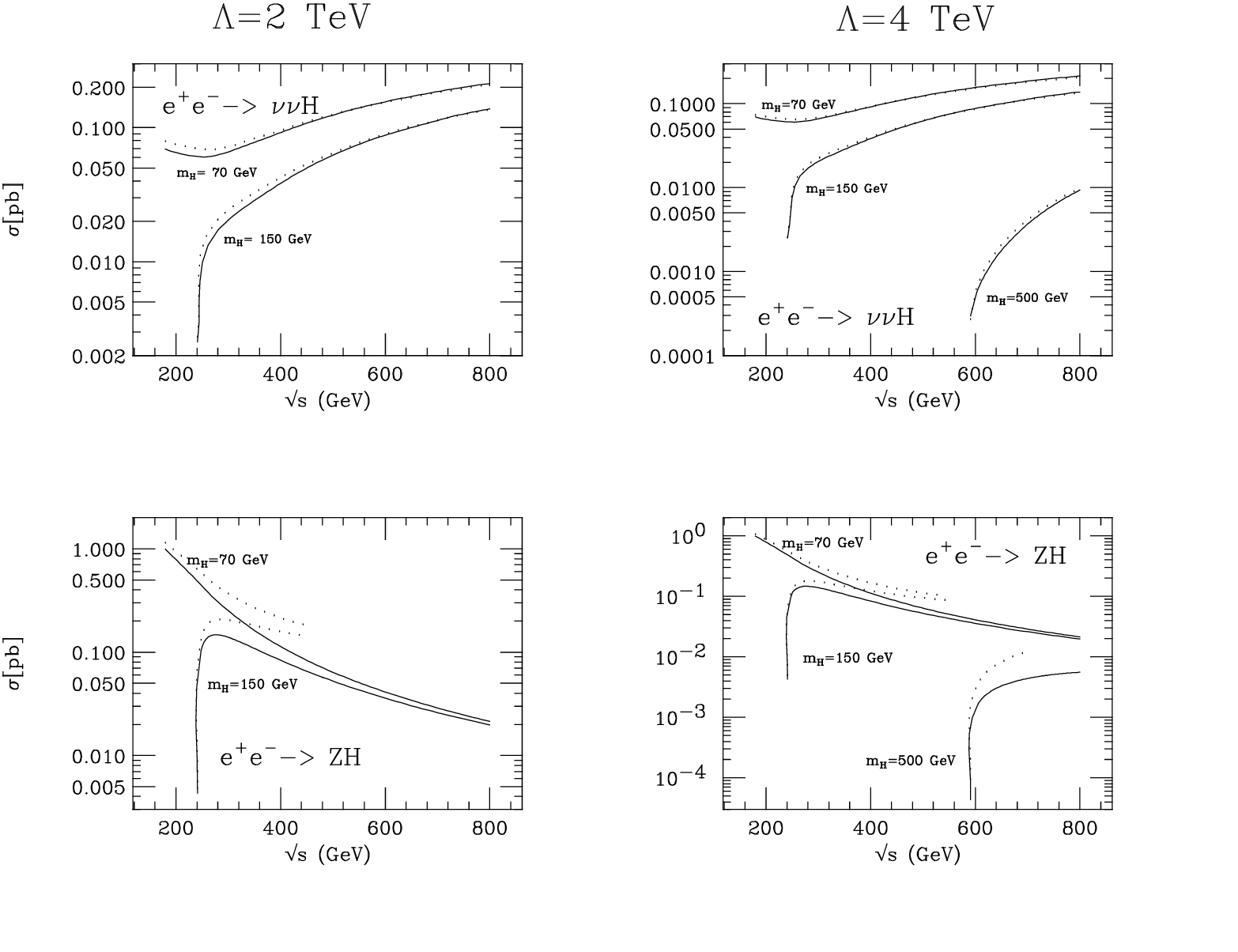}}
\centerline{\box2}

\vskip 130pt
\thecaption{Fig. 3}{Total cross sections for the process $ e^+ e^-
\rightarrow H \bar \nu \nu $ and $ e^+ e^- \rightarrow Z H $. The values
for the constants $ \alpha_i $ and other restrictions are specified in the
text. The solid lines correspond to the \sm\ predictions whereas the dotted
ones represent the effective lagrangian results.}
\bigskip\bigskip

In evaluating the cross section we have expressed the weak-mixing angle,
the \vev, etc. in terms of the above-mentioned input parameters. The
expressions are cumbersome and will not be given explicitly here; they
can be extracted from the results presented in Ref.~\cite{bw}.

It is clear from the plots in figure 3 that the new physics
contributions to the $ e^+ e^- \rightarrow H \bar \nu \nu $ process are
unobservable. In contrast the deviations from the \sm\ for the reaction
$ e^+ e^- \rightarrow Z H $ can be quite significant. To illustrate
the implications of this result we calculated the statistical significance
$ \nsd $ of the deviations from the \sm\ for the Bjorken process.
This quantity is defined by \begin{equation}
\nsd = {\left| \sigma - \sigma\lowti{ SM } \right| L \over \sqrt {\sigma L \ }
}
\end{equation}
where $ \sigma $ and $ \sigma\lowti{ SM } $ denote the total and
\sm\ cross sections respectively, and $L$ denotes the luminosity. The
results are presented in figure 4.

\bigskip\bigskip
\setbox2=\vbox to 100 pt {\epsfxsize=5 truein\epsfbox[0 -50 612
642]{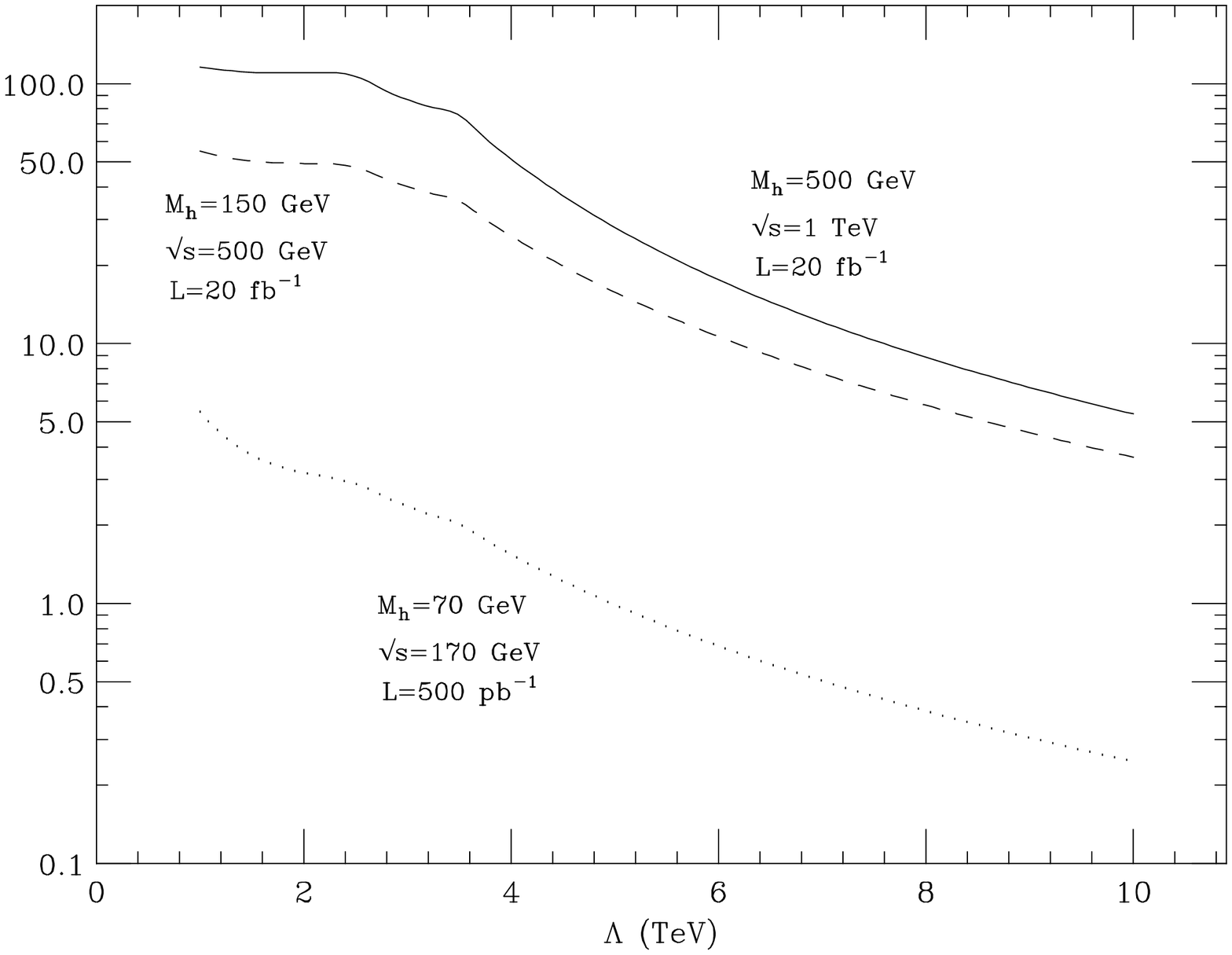}}
\centerline{\box2}

\vskip 140pt

\thecaption{Fig. 4}{Statistical significance of the new physics effects
for the process $ e^+ e^- \rightarrow Z H $.
The non-monotonic behavior of the curves is due to our having chosen
the $ \alpha_i $ which saturate the exisiting bounds
(at the $ 3 \sigma $ level)
obtained in (\ref{eq:lepbounds},\ref{eq:tbounds}), see the text for details.}
\bigskip\bigskip

As above, we have chosen the $ \alpha_i $ which saturate the bounds
(\ref{eq:lepbounds},\ref{eq:tbounds}) at the $ 3 \sigma $ level. The results in
figure 4
then give the {\it maximum possible value of $ \nsd $ to be observed at
LEP2} which is also consistent with the exisiting measurements.
As can be seen from this figure, the sensitivity of LEP2 reaches
several \tev\ when this process is considered. The situation is further
improved for a Next Linear Collider of $ 500 \gev $ CM energy and $ 20
fb^{-1}$ luminosity.

Finally we compute the sensitivity to the scale of new physics $ \Lambda
$ as a function of the energy of the $ e^+ e^- $ collider used to probe
this type of new physics. We present the various discovery limit
contours in figure 5.


\setbox2=\vbox to 270 pt {\epsfxsize=5 truein\epsfbox[0 -250 612
442]{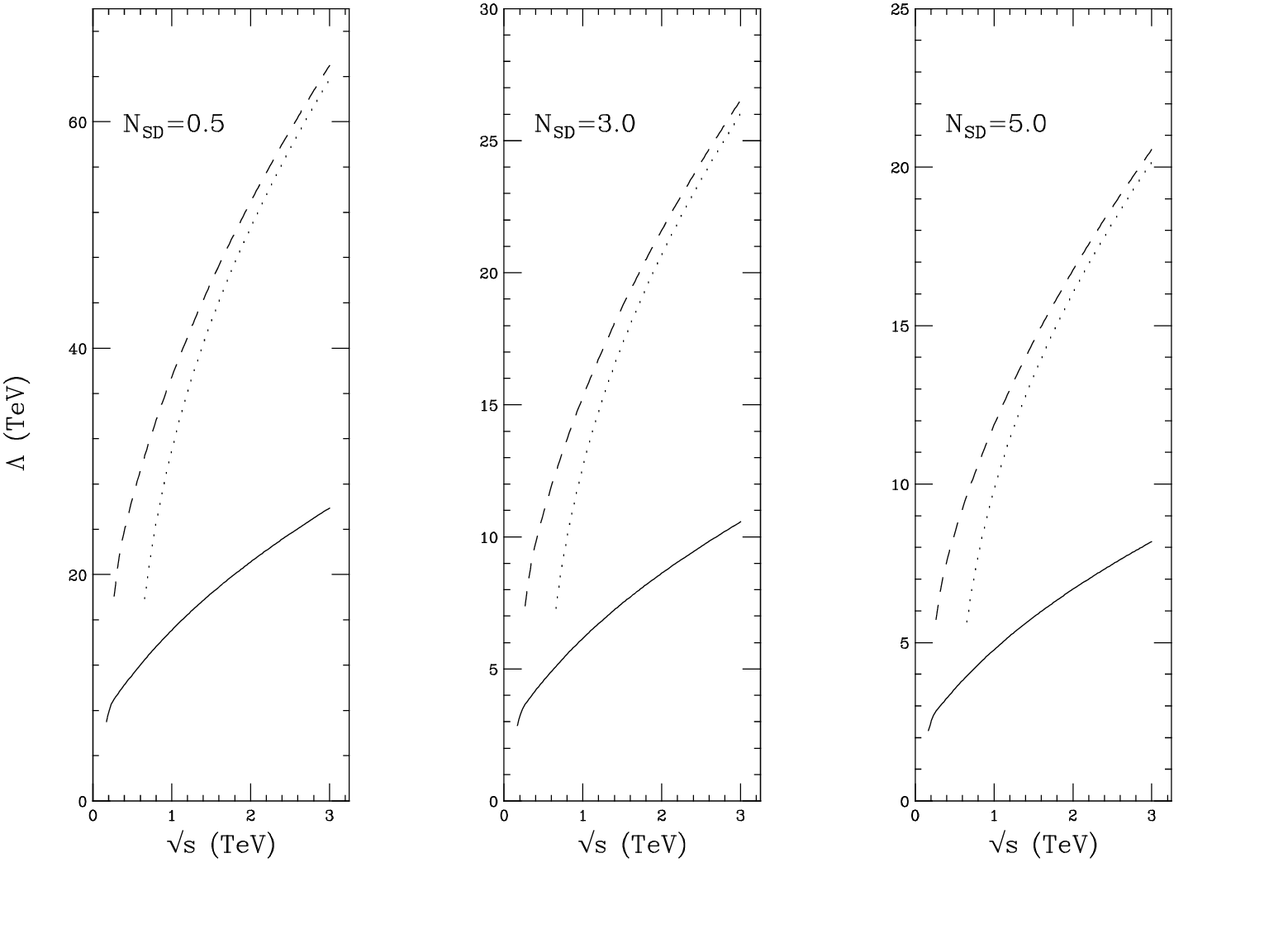}}
\centerline{\box2}

\bigskip\bigskip

\thecaption{Fig. 5}{Sensitivity to the scale of new physics $ \Lambda $
for a given CM energy $ \sqrt{ s } $ using the reaction
$ e^+ e^- \rightarrow Z H $. Denoting by $L$ the luminosity, the
curves correspond to
$ \mh=70 \gev,  \ L = 0.5 \; \hbox{fb}^{ - 1 } $: solid line;
$ \mh=150 \gev, \ L = 20  \; \hbox{fb}^{ - 1 } $: dashed line;
$ \mh=500 \gev, \ L = 20  \; \hbox{fb}^{ - 1 } $: dotted line.
We chose $ \alpha_i = 1 $ (for which the current $3 \sigma $
bounds imply $ \Lambda \gesim 2.7 \tev $).}

\bigskip\bigskip

The above results demonstrate a significant sensitivity of near-future
accelerators to new physics effects. The approach presented here,
applicable to the case where there is a light Higgs excitation, is based
on the segregation of those new-physics effects that can occur at tree
level. In case  the careful examination of this process (and others
similar to it) provides no hint of deviations from the \sm\ it will be
necessary to conclude that the corresponding operators are suppressed in
the underlying theory. This is a non-trivial statement that will
eliminate several types of interactions. For example, the absence of a
deviations in the $ e e Z $ vertex significantly constrains the mixings
and masses of a new neutral vector boson and also the mixings of the
electron with new heavy leptons. Since the general form of the
interactions that give rise to the operators (\ref{eq:operators}) is
known~\cite{Arzt.et.al.}, even the absence of deviations from the \sm\
can be translated into useful information: the various contributions to
the effective operators must either be suppressed or cancelations must
be present.

It is interesting to provide some models which can generate the
operators studied in this paper. One can consider, for example the
addition to the \sm\ of a vector-like fermion $ \Psi^I $ which is an
$ \su2_L$
triplet and a $ \ui_Y$ singlet; in this case $ \alpha_{ \phi \ell }\up{
1, 3 } $ will be non-zero and proportional to the $ \bar \ell \tau^I \Psi^I
\tilde \phi $ coupling constant. The same operators are
generated by a heavy $Z'$ exchange; in this case $ \Lambda $ corresponds
to the \vev\ in the new $ \ui $ and the \sm\ scalar doublet and leptons
are assumed to have non-zero quantum numbers in the new $ \ui $ group.
We also note that the MSSM \cite{Gunion/Haber} does not generate the
operators under consideration (unless R-parity violations are included).
The observation of strong deviations from the \sm\ in the above processes
would then argue against this model.

It is worth noticing that  at high energy colliders the $H \bar \nu \nu$
production channel, having much bigger cross section than the $ZH$ one,
can be used to measure $H$ properties (e.g. $m_H$) while the $ZH$ should
be utilized to observe
deviations from the \sm\ prediction. Therefore the both processes considered
here would be needed and are in fact complementary: $H \bar \nu \nu$ having
very small corrections, but providing large production rate and
$ZH$ with it's substantial sensitivity to non-standard physics.

\bigskip\bigskip

{\large \bf Acknowledgments}

One of the authors (B.G.) expresses his thanks for a warm hospitality
to the Physics Department at University of California,  Riverside,
where this research has been performed. Contribution of B.G. was partially
supported by the Committee for Scientific Research under grant BST-475.
Part of this work was supported through funds provided by the Department
of Energy under contract DE-FG03-94ER40837 and by
the University of California under the AAFDA program.

\end{document}